\documentclass[prb,twocolumn,aps,superscriptaddress,showpacs,floatfix]{revtex4}

\usepackage{amsmath}
\usepackage{amssymb}
\usepackage{amsbsy}
\usepackage{graphicx}
\usepackage{color}
\usepackage{epstopdf}
\usepackage{mathrsfs}

\begin{document}
	
	\title{Dynamical evolution in a one-dimensional incommensurate lattice with $\mathcal{PT}$ symmetry}
	
	\author{Zhihao Xu}
	\email{xuzhihao@sxu.edu.cn}
	\affiliation{Institute of Theoretical Physics and State Key Laboratory of Quantum Optics and Quantum Optics Devices, Shanxi University, Taiyuan 030006, China}
	\affiliation{Beijing National Laboratory for Condensed Matter Physics, Institute of Physics, Chinese Academy of Sciences, Beijing 100190, China}
	\affiliation{Collaborative Innovation Center of Extreme Optics, Shanxi University, Taiyuan 030006, P.R.China}
	
	\author{Shu Chen}
	\affiliation{Beijing National Laboratory for Condensed Matter Physics, Institute of Physics, Chinese Academy of Sciences, Beijing 100190, China}
	\affiliation{School of Physical Sciences, University of Chinese Academy of Sciences, Beijing, 100049, China}
	\affiliation{Yangtze River Delta Physics Research Center, Liyang, Jiangsu 213300, China}
	
	\begin{abstract}
		We investigate the dynamical evolution of a parity-time ($\mathcal{PT}$) symmetric extension of the Aubry-Andr\'{e} (AA) model, which exhibits the coincidence of a localization-delocalization transition point with a $\mathcal{PT}$ symmetry breaking point. One can apply the evolution of the profile of the wave packet and the long-time survival probability to distinguish the localization regimes in the $\mathcal{PT}$ symmetric AA model. The results of the mean displacement show that when the system is in the $\mathcal{PT}$ symmetry unbroken regime, the wave-packet spreading is ballistic, which is different from that in the $\mathcal{PT}$ symmetry broken regime. Furthermore, we discuss the distinctive features of the Loschmidt echo with the post-quench parameter being localized in different $\mathcal{PT}$ symmetric regimes.
		
	\end{abstract}

	\maketitle
	\section{Introduction}
	
	Exploration localization induced by disorder is a long-standing research topic in condensed-matter physics. The Anderson localization induced by random disorder first proposed by Anderson \cite{Anderson} has found its way across a wide range of different fields, such as cold atomic gases\cite{Billy,Roati,Kondov,Jendrzejewski,Semeghini,Pasek,Hainaut,Richard}, quantum optics\cite{Sperling,Wiersma,Aegerter,Schwartz,Lahini}, acoustic waves\cite{Strybulevych} and electronic systems\cite{Katsumoto}. In comparison with the random disorder cases, the quasicrystal systems constitute an intermediate phase between periodic lattices and disorder media, which display long-range order but no periodicity. A paradigmatic example of a one-dimensional quasicrystal system is the Aubry-Andr\'{e} (AA) model, which has attracted increasing interest in recent years \cite{Aubry,Thouless,Sokoloff,Hofstadter,CRDean,Grempel,Kohmoto,SDasSarma,Lahini1,Biddle,Biddle1,Pouranvari,Aulbach,Modugno,Larcher,Ingold,Kraus,Lang,Silva,Zhihao,Zhihao1,Shiliang,Basko,Schreiber,Iyer,Bordia}. A typical feature of the AA model is that, the system undergoes a metal-insulator transition when the amplitude of the quasicrystal potential exceeds a finite critical value, which is determined by the self-duality property \cite{Thouless}. The AA model has been experimentally realized by the cold atomic technique in bichromatic optical lattices \cite{Roati}. 
	
	On the other hand, thanks to the impressive progress in controlling quantum matter in recent decades, the realizations of the real-time dynamics of quantum systems has been achieved on various experimental platforms, such as ultra-cold atoms in optical lattices, trapped ions, and photonic lattices, and the inaccessible dynamical phenomena have been researched. In particular, the dynamical evolution of a wave packet in a disordered system has drawn considerable interest. One tries to understand the relation between the energy spectrum and the dynamical propagation of the wave packet \cite{Kohmoto1,Ostlund,Kohmoto2,Kohmoto3,Abe,Katsanos,Geisel,Ketzmerick,Huckestein,Ketzmerick1,ZhenjunZhang,Dadras,Sinha,Santos,Silberstein}. The dynamical observation of wave-packet evolution and many-body localization in one-dimensional incommensurate optical lattices has also been reported in recent works \cite{Luschen,Kohlert,Zhihao2}.  The dynamical phase transition based on the Loschmidt echo is another topic attracting wide attention \cite{Heyl,Karrasch,Canovi,Jalabert,Cucchietti,Gorin,HTQuan,Jafari,Budich,Vajna,Vajna1,Sharma,Bhattacharya,YangChao,Kennes,Szpak,Jurcevic,Pagano,Vogel,XYGuo,KWang,TTian,KXu,HYin,TongLiu,YanxiaLiu,Peotta,XTong}. A dynamical phase transition occurs, when the quench process goes across the critical point. It corresponds to the vanishing of the Loschmidt echo, which has been successfully applied in the AA model \cite{YangChao} and its extensions \cite{Zhihao2,YanxiaLiu,XTong}.
	
	Recently, great interest has been devoted to the interplay of non-Hermiticity and disorder, which brings a new perspective of the localization properties \cite{YanxiaLiu,Hatano1,Hatano2,Kolesnikov,ZhongPingGong,Tzortzakakis,YHuang1,YHuang2,QBZeng1,QBZeng2,TongLiu1,YanxiaLiu1,YanxiaLiu2,YanxiaLiu3,Longhi1,Longhi2,Longhi3,CHLiang,Cortes,Harter,Rivolta,HuiJiang,DWZhang,Claes,LJZhai,Okuma,Tzortzakakis1,CWang1,Hamazaki,CHLiu,Tzortzakakis2,Yusipov1,Balasubrahmaniyam,Goldsheid,Molinari,Markum,Chalker}. Non-Hermitian models are found in open systems exchanging energy or particles with the environment. For a non-Hermitian system, the non-Hermiticity is generally obtained by introducing nonreciprocal hopping terms or gain and loss potentials. According to the random matrix theory, the spectral statistics of non-Hermitian disorder systems exhibits different properties from the Hermitian ones \cite{Goldsheid,Molinari,Markum,Chalker}. The Hatano-Nelson model describing the interplay of the nonreciprocal hopping and random disorder exhibits a finite localization-delocalization transition \cite{Hatano1,Hatano2,Kolesnikov,ZhongPingGong}. Non-Hermitian extensions of the AA models realized by introducing non-reciprocal hopping or $\mathcal{PT}$ symmetric potential have been investigated in Refs. \cite{YanxiaLiu,QBZeng1,QBZeng2,TongLiu1,YanxiaLiu1,YanxiaLiu2,YanxiaLiu3,Longhi1,Longhi2,Longhi3,HuiJiang}. For a $\mathcal{PT}$ symmetric extension of the AA model, one can find the coincidence of a localization transition point with a $\mathcal{PT}$ symmetry breaking point by both numerical and analytical calculations \cite{Longhi1,Longhi2,HuiJiang}. The analytical results show that the localization length in the insulator phase is independent of energy which is similar to the Hermitian AA case. On the other hand, the energy spectrum is gapless in the metallic phase which is unlike the Hermitian AA model \cite{Longhi2}. Due to the anomalous energy spectrum of the $\mathcal{PT}$ symmetric AA model, some interesting questions arise here: What are the features of the dynamical evolution of the wave packet in the one-dimensional incommensurate lattice with $\mathcal{PT}$ symmetry? Can the Loschmit echo method be applied to detect the localization transition for a $\mathcal{PT}$ symmetric AA model?
	
	In this work, to address these questions, we study the dynamical evolution of the $\mathcal{PT}$ symmetric AA model in different localization regimes by applying the propagation of the profile of the wave packet, the long-time survival probability, the evolution of the mean displacement, and the Loschmidt echo dynamics. As a comparison, we also discuss the case of the standard AA model. We find that the propagation of the profile of the wave packet and the long-time survival probability exhibit distinctive features in different localization regimes for both the Hermitian and non-Hermitian cases . The behaviors of the evolution of the mean displacement are dependent on the breaking of the $\mathcal{PT}$ symmetry in the non-Hermitian AA model, while for the Hermitian case, it displays different diffusion exponents in different localization regimes. The Loschmidt echo dynamics with the post-quench parameter localized in the $\mathcal{PT}$ symmetry unbroken regime can be used to detect the localization transition, which is similar to the Hermitian case. However, for the post-quench parameter localized in the $\mathcal{PT}$ symmetry broken regime, our results indicate that the detection of the dynamical phase transition is unavailable using Loschmidt echo dynamics.
	
	
	\section{Model and Hamiltonian}
	\begin{figure}[tbp]
		\begin{center}
			\includegraphics[width=.5 \textwidth] {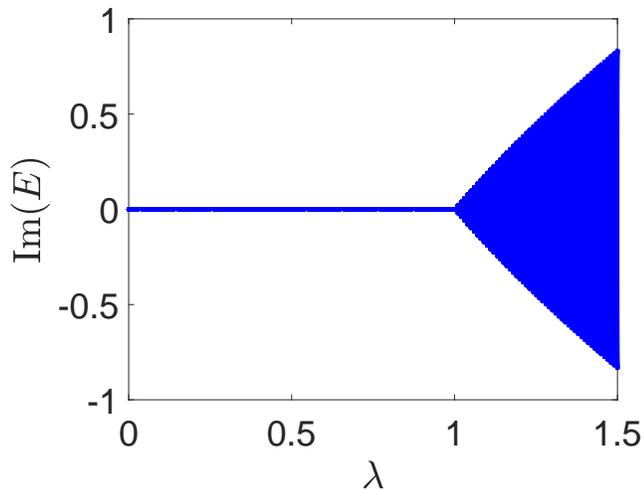}
		\end{center}
		\caption{(Color online) Imaginary part of the energy spectrum $\mathrm{Im}(E)$ for the $\mathcal{PT}$ symmetric AA model as a function of the modulation strength $\lambda$. Here, $J=1$ and $L=377$.}\label{Fig11}
	\end{figure}

	\begin{figure}[tbp]
		\begin{center}
			\includegraphics[width=.5 \textwidth] {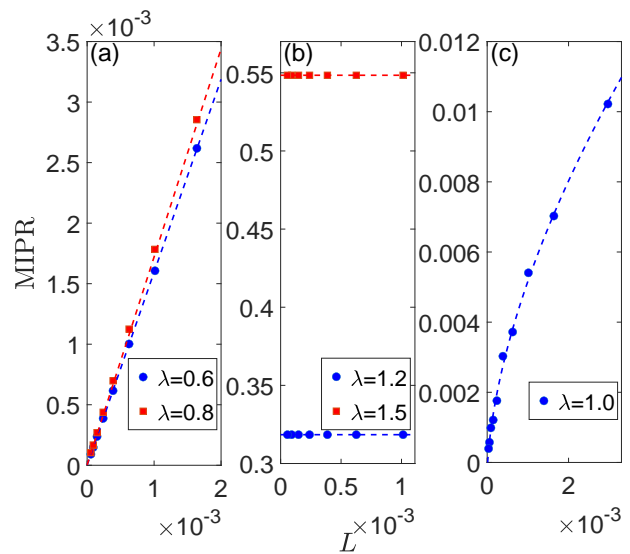}
		\end{center}
		\caption{(Color online) Scaling of MIPR for the $\mathcal{PT}$ symmetric AA model in (a) $\lambda<J$, (b) $\lambda>J$, and (c) $\lambda=J$, respectively. The dashed line indicates a power-law fitting. Here, $J=1$.}\label{Fig1}
	\end{figure}
	
	We consider a non-Hermitian extension of the AA model with a complex incommensurate lattice subjected to $\mathcal{PT}$ symmetry, which can be described by the Hamiltonian
	\begin{equation}\label{eq1}
		\hat{H}=-\sum_j J(\hat{c}_j^{\dagger}\hat{c}_{j+1}+H.c.)+\sum_j \lambda_j\hat{n}_j,
	\end{equation}
	where $\hat{c}_j$ is the annihilation operator of fermions at the $j$-th site, and $J$ denotes the strength of the hopping term. The on-site $\mathcal{PT}$ symmetric potential can be written as
	\begin{equation}\label{seq1}
		\lambda_j=\lambda \exp{(-i2\pi \alpha j)}
	\end{equation}
	with the amplitude $\lambda$ and $\alpha$ being irrational, which leads to the non-Hermiticity of the Hamiltonian (\ref{eq1}). The potential is the complexification of the standard AA model with on-site potential $\lambda_j=2\lambda^{\prime}\cos{(2\pi \alpha j +\varphi)}$ taking $\varphi=i h$, and the limits $\lambda^{\prime}\to 0$, $h\to \infty$, keeping $\lambda^{\prime}\exp{(h)}=\lambda$ finite. It can be realized in the silicon waveguide platform, the cold atomic gases, and ion chains \cite{Bylinskii1,Bylinskii2,Bonetti,Benassi,Mandelli,Kiethe}. Numerical and analytical results exhibit a metal-insulator phase transition at $\lambda=J$ for the non-Hermitian system, which also corresponds to $\mathcal{PT}$ symmetric breaking \cite{Longhi1,Longhi2}. Figure \ref{Fig11} shows the imaginary part of the energy spectrum $\mathrm{Im}(E)$ of the $\mathcal{PT}$ symmetric AA model as a function of $\lambda$. As seen in Fig. \ref{Fig11},  when $\lambda<J$, all the energy spectra are real values, while for $\lambda>J$ the energies become complex. 
	
	To characterize the localized properties of the $\mathcal{PT}$ symmetric AA model, we can investigate the fractal dimension of the wave functions $\beta$ defined by $\mathrm{MIPR} \propto L^{-\beta}$, where the mean inverse participation ratio (MIPR)
	\begin{equation}\label{eq2}
		\mathrm{MIPR}=\frac{1}{L}\sum_{n=1}^{L}\sum_{j=1}^{L}\frac{|\psi_{n,j}|^4}{|\psi_{n,j}|^2}
	\end{equation}
	with $\psi_{n,j}$ being the amplitude of the eigenstate $|\psi_{n}\rangle$ of the eigenvalue $E_n$ at the $j$th site and $L$ is the size of the lattice. It is known that $\beta=0$ for the localized regime, $\beta=1$ for the extended regime, and $0<\beta<1$ corresponding to a multifractal phase. Figure \ref{Fig1} shows the scaling of the MIPR for the $\mathcal{PT}$ symmetric AA model in different localized regimes. As shown in Fig. \ref{Fig1}(a) for $\lambda/J=0.6$ and $0.8$ in the extended regime, $\beta=1$ and the MIPRs tend to $0$ with the increase of $L$. In the localized regime, taking $\lambda/J=1.2$ and $1.5$ as examples shown in Fig. \ref{Fig1}(b), the MIPRs are finite and independent of $L$ with $\beta=0$. For $\lambda/J=1$ [Fig. \ref{Fig1}(c)], when $L\to \infty$, the MIPR approaches $0$ with $\beta\approx 0.62$. This implies that the non-Hermitian AA model is localized in the multifractal phase for $\lambda/J=1$. As a comparison, we briefly recall the main conclusions for the Hermitian AA model, corresponding to $\lambda_j=2\lambda^{\prime}\cos{(2\pi \alpha j)}$, with $\alpha$ being irrational. The system exhibits a transition from the delocalized phase for $\lambda^{\prime}<J$ with $\beta=0$ to localized region for $\lambda^{\prime}>J$ with $\beta=1$, and for $\lambda^{\prime}=J$, the system is localized in the multifractal phase with $\beta=0.5$ \cite{Zhihao2,Geisel}. We can see that the $\mathcal{PT}$ symmetric AA model exhibits similar localization properties to those of the Hermitian one. However, their dynamical evolutions of both cases show some distinctive behaviors.
	
	In this paper, we study the dynamical behaviors of a non-Hermitian AA model with $\mathcal{PT}$ symmetric potentials described by the Hamiltonian (\ref{eq1}) in real space on a ring, and we take $J=1$ as the energy unit. The irrational number $\alpha=(\sqrt{5}-1)/2$ is chosen of which the approximants are $F_{\mu-1}/F_{\mu}$, with $F_{\mu}$ being the $\mu$th Fibonacci number defined by $F_{\mu}=F_{\mu-1}+F_{\mu-2}$, with $F_0=F_1=1$ yielding the size of the lattice $L=F_{\mu}$.
	
	
	\section{Wave packet dynamics}
	
	\begin{figure}[tbp]
		\begin{center}
			\includegraphics[width=.5 \textwidth] {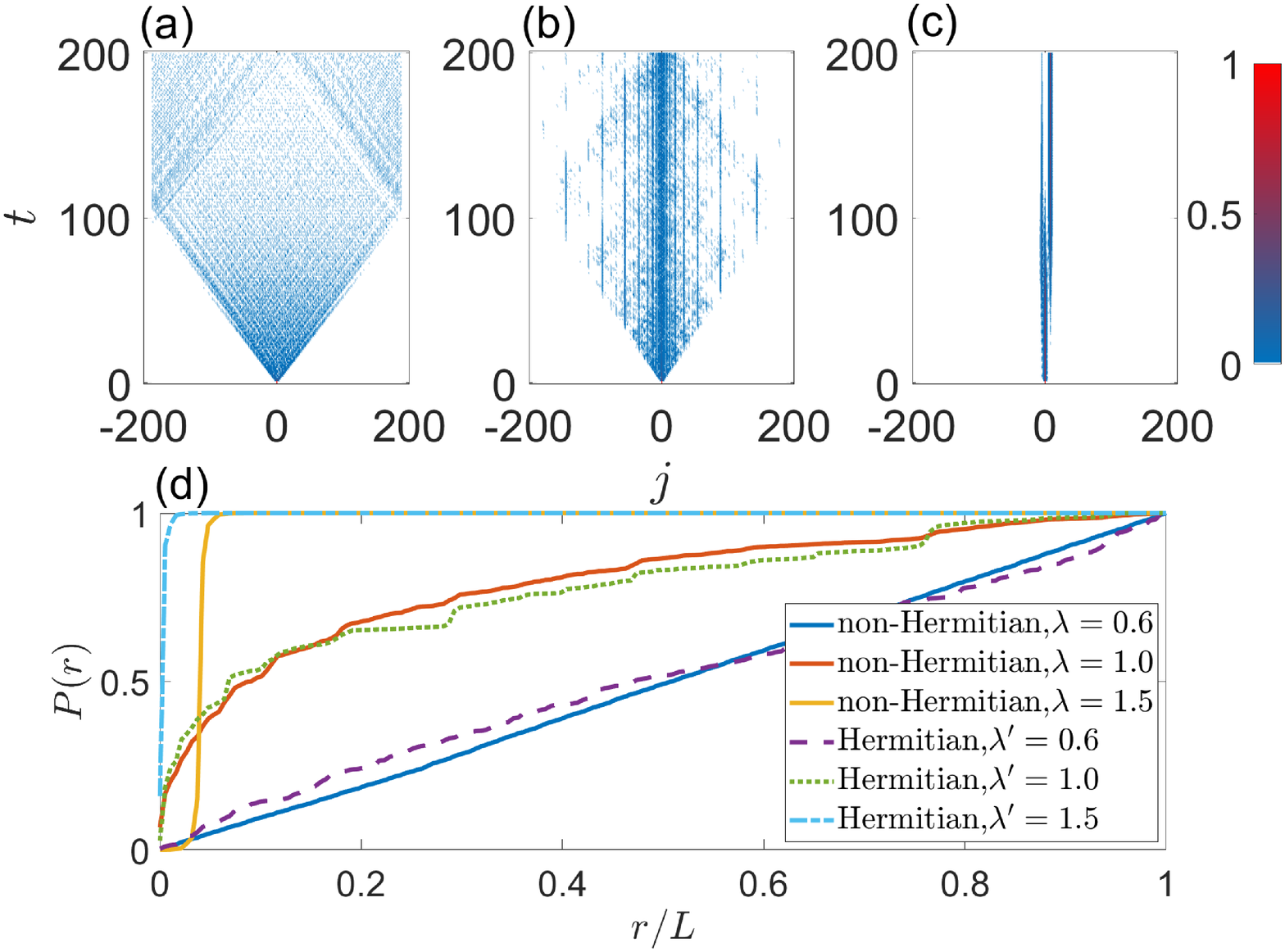}
		\end{center}
		\caption{(Color online) (a)-(c) Evolutions of the profile of the wave function $\rho_j(t)$ with the non-Hermitian modulation amplitude (a) $\lambda=0.6$, (b) $\lambda=1.0$, and (c) $\lambda=1.5$, respectively. (d) Long-time survival probability $P(r)$ for both the non-Hermitian and the Hermitian cases.  For the non-Hermitian cases, we choose  $\lambda=0.6$ and $1.0$ at $t=800$ and $\lambda=1.5$ at $t=400$. For the Hermitian cases, we choose the Hermitian modulation amplitude $\lambda^{\prime}=0.6$, $1.0$, and $1.5$ at $t=8000$. Here, $J=1$ and $L=377$. }\label{Fig2}
	\end{figure}
	
	We first investigate the expansion dynamics of the wave packet $|\Psi(0)\rangle=|j_0\rangle$ initially localized at the center of the lattice $j_0$ governed by the non-Hermitian AA model described by the Hamiltonian (\ref{eq1}). The evolution wave function at time $t$ can be written as
	\begin{equation}\label{eq3}
		|\Psi(t)\rangle =\frac{1}{\sqrt{\mathcal{N}}}e^{-i\hat{H}t}|\Psi(0)\rangle,
	\end{equation}
	with $\mathcal{N}$ being the normalization coefficient of the time-evolution wave function $|\Psi(t)\rangle$. Here, we use a normalized wave function in our $\mathcal{PT}$ symmetric model to avoid an exponential rise of the wave function with time in the $\mathcal{PT}$ symmetry broken regime. For a given initial state $|\Psi(0)\rangle$, to numerically calculate the wave function $|\Psi(t)\rangle$ at time $t$, we divide time $t$ into $\bar{M}$ intervals with $dt=t/\bar{M}$. In the limit $\bar{M}\to \infty$, $dt \to 0$, and drop the $O(dt^2)$ terms. The wave function at time $(\bar{m}+1)dt$	is given by
	\begin{equation}
		|\Psi((\bar{m}+1)dt)\rangle =\frac{(1-iHdt)|\Psi(\bar{m}dt)\rangle}{\langle \Psi(\bar{m}dt)| (1+iHdt)(1-iHdt)|\Psi(\bar{m}dt)\rangle}, \notag
	\end{equation}
	with
	\begin{equation}
		|\Psi(dt)\rangle =\frac{(1-iHdt)|\Psi(0)\rangle}{\langle \Psi(0)| (1+iHdt)(1-iHdt)|\Psi(0)\rangle}. \notag
	\end{equation}
	and $\bar{m}$ is the iteration times. After the $\bar{M}$ times iteration processes, we can obtain $|\Psi(t)\rangle$. 
	
	To obtain an intuitive picture, we study the profile of the wave function at time $t$ given by
	$\rho_{j}(t)=\langle \Psi_j(t)|\Psi_j(t)\rangle$ shown in Figs. \ref{Fig2}(a)-(c) for different $\lambda$ with $L=F_{13}=377$. For the extended case [Fig. \ref{Fig2}(a) with $\lambda=0.6$], the initial state localized at the center of the lattice rapidly expands to the whole system, and after some time intervals, the profile of the wave function exhibits an extended character. In the multifractal phase,  which is shown in Fig. \ref{Fig2}(b), the center part of the profile of the wave function decays with time and keeps finite in our observation time. The expanding part seems to show a multifractal property with time. When we turn the strength of quasi-periodic modulation $\lambda$ to the localized regime, e.g., $\lambda=1.5$ in Fig. \ref{Fig2}(c), the wave function exhibits non-Hermitian jumps between distant sites, which only occurs in the localized regime of the non-Hermitian system and does not have a Hermitian analog \cite{Wiersma,Tzortzakakis2,Yusipov1,Balasubrahmaniyam}. 
	
	To further observe the dynamical behaviors of the non-Hermitian system in different phases, we define the long-time survival probability 
	\begin{equation}\label{eq4}
		P(r)=\lim_{t\to\infty}\sum_{|j-j_0|\le r/2}\langle \Psi_j(t)|\Psi_j(t)\rangle,
	\end{equation}
	which represents the normalized probability of detecting the wave packet in sites within the region $(-r/2,r/2)$ in the long-time limit \cite{Santos}. The recent works for random matrix models show that $P(r)$ provides crucial information about localization properties \cite{Kravtsov,Herrera,Tomasi}. Figure \ref{Fig2}(d) shows $P(r)$ as a function of $r/L$ for both non-Hermitian and Hermitian cases with $L=F_{13}=377$ in the long-time limit. When the strength of the modulation is localized in the extended regime, $P(r)$ of both cases increase linearly with $r$, \emph{i.e.}, $P(r)\approx r/L$, since the probability of finding the normalized wave packet at each site is the same for both cases. For $\lambda=\lambda^{\prime}=1$, $P(r)$ of both cases show the multifractal feature with $P(r) \propto (r/L)^{\tilde{\beta}}$ and the power-law exponent $\tilde{\beta}\in(0,1)$. For the Hermitian case in the localized regime with $\lambda^{\prime}=1.5$ shown in Fig. \ref{Fig2}(d), we can see that $P(r)$ is finite at $r=0$, and it presents an exponential rise, and rapidly reaches $(r/L)^0=1$. However, for the non-Hermitian case with $\lambda=1.5$, the value of $P(r)$ approaches $0$ in the $r/L\ll 1$ limit, and it exhibits an exponential increase after some $r$, and rapidly reaches $(r/L)^0=1$. According to our results, the long-time survival probability $P(r)$ is proportional to $(r/L)^{\tilde{\beta}}$, and it can be applied to the non-Hermitian case to distinguish the long-time dynamical behaviors of the wave packet in different localization regimes. We also find that $P(r)$ of the non-Hermitian case in the localized regime presents a vacuum space in the small-$r$ limit that is different from the Hermitian one.
	
	\begin{figure}[tbp]
		\begin{center}
			\includegraphics[width=.5 \textwidth] {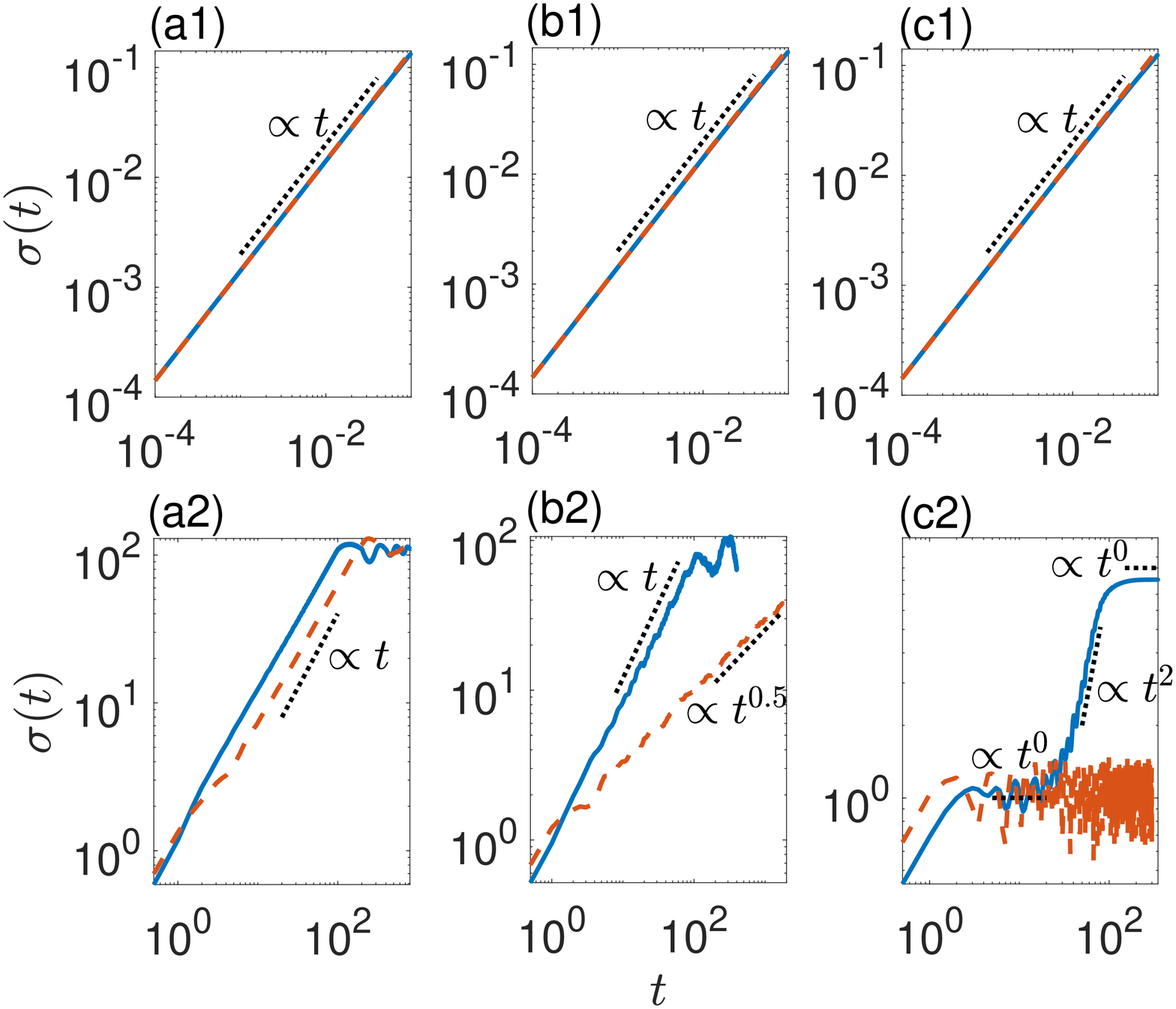}
		\end{center}
		\caption{(Color online) Time evolution of the mean displacement $\sigma(t)$ for the non-Hermitian cases marked by the solid lines and the Hermitian cases marked by the dashed lines. The black dotted lines represent the power-law fitting curves. (a1) and (a2) are for $\lambda=\lambda^{\prime}=0.6$. (b1) and (b2) are for $\lambda=\lambda^{\prime}=1.0$. (c1) and (c2) are for $\lambda=\lambda^{\prime}=1.5$. $\lambda$ and $\lambda^{\prime}$ are the non-Hermitian and Hermitian modulation amplitudes, respectively. The top row corresponds to the evolution in the short-time stage, and the bottom row exhibits the long-time evolution processes. Here, $J=1$ and $L=377$. }\label{Fig3}
	\end{figure}
	
\begin{figure}[tbp]
		\begin{center}
			\includegraphics[width=.5 \textwidth] {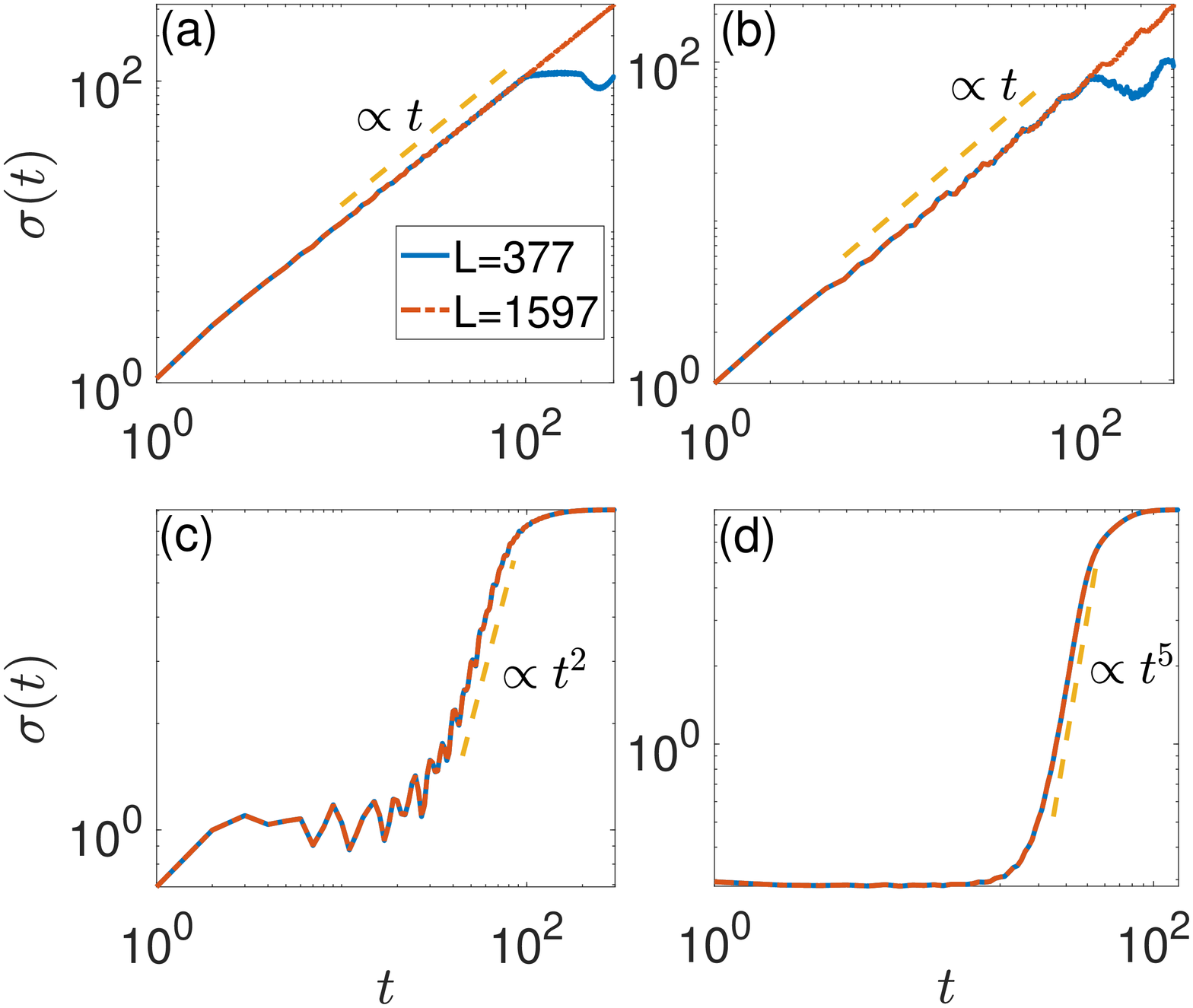}
		\end{center}
		\caption{(Color online) Time evolution of the mean displacement $\sigma(t)$ with different $L$ and the non-Hermitian modulation strength (a) $\lambda=0.8$, (b) $\lambda=1.0$, (c) $\lambda=1.5$, and (d) $\lambda=3$, respectively. Here, $J=1$. }\label{Fig31}
	\end{figure}

	The wave-packet spreading dynamics can be described by the evolution of the mean displacement \cite{ZhenjunZhang,CMDai} given by
	\begin{equation}\label{eq5}
		\sigma(t)=\sqrt{\sum_{j=1}^{L}\langle \Psi_j(t)|(j-j_0)^2|\Psi_j(t)\rangle}.
	\end{equation}
	In general, for a long expansion time in a Hermitian case, the mean displacement obeys the power law $\sigma(t)\sim t^{\tilde{\gamma}}$. Apart from ballistic spread for $\tilde{\gamma}=1$ and localization or diffusive transport for $\tilde{\gamma}=0$ or $1/2$, subdiffusion for $0<\tilde{\gamma}<1/2$ and superdiffusion for $1/2<\tilde{\gamma}<1$ can occur in some quasi-periodic lattices. When $\tilde{\gamma}>1$, the initial localized wave packet exhibits a hyperdiffusion transport for a certain time scale. Figure \ref{Fig3} shows the time evolution of $\sigma(t)$ with an initial localized wave packet $|j_0\rangle$ at the center of the lattice $j_0$. The evolution of $\sigma(t)$ in different localization regimes for both the non-Hermitian and Hermitian cases is marked by solid lines and dashed lines, respectively. During the first stage of evolution shown in Figs. \ref{Fig3}(a1)-\ref{Fig3}(c1) with $\lambda=\lambda^{\prime}=0.6$, $1.0$, and $1.5$, respectively, the wave packet spreading is ballistic with the same power-law indices $\tilde{\gamma}=1$ independent of the hermiticity of the system and the strength of the modulation. The long-time evolution of $\sigma(t)$ is shown in Figs. \ref{Fig3}(a2)-\ref{Fig3}(c2). During the second stage, as shown in Fig. \ref{Fig3}(a2) with $\lambda=\lambda^{\prime}=0.6$, both cases present ballistic diffusion with $\tilde{\gamma}=1$. When the on-site potential amplitude is localized in a multifractal regime, the diffusion exponent $\tilde{\gamma}=0.5$ for the Hermitian case, while it still exhibits a ballistic diffusion ($\tilde{\gamma}=1$) for the non-Hermitian case with $\lambda=1$. For $\lambda=\lambda^{\prime}>1$ cases [$\lambda=\lambda^{\prime}=1.5$ shown in Fig. \ref{Fig3}(c2)],  after the first stage spreading, $\sigma(t)$ exhibits an oscillating characteristic and the diffusion exponent $\tilde{\gamma}=0$. However, the non-Hermitian case is quite different. After the ballistic spreading stage, $\sigma(t)$ enters into a temporary stage whose behavior is similar to that in the second stage of the Hermitian case. It then presents a hyperdiffusion with $\tilde{\gamma}\approx 2$ in a short-time interval, which corresponds to a non-Hermitian jump process \cite{Wiersma,Tzortzakakis2,Yusipov1,Balasubrahmaniyam}. The diffusion exponent $\tilde{\gamma}>1$ in this stage is dependent on the value of $\lambda$ [see also Fig. \ref{Fig31}(d)]. Finally, the wave packet seems to be frozen with $\tilde{\gamma}=0$. According to our results, we can see that the evolution of the mean displacement $\sigma(t)$ shows a ballistic spreading for the non-Hermitian AA model (\ref{eq1}) in the $\mathcal{PT}$ symmetry unbroken regime independent of its localization properties, which is different from the Hermitian case. However, when the non-Hermitian AA model enters into the localized regime corresponding to the $\mathcal{PT}$ symmetry broken regime, $\sigma(t)$ successively undergoes ballistic diffusion to localization to hyperdiffusion and back to localization. The results of $\sigma(t)$ suggest that the dynamics in the $\mathcal{PT}$ symmetry unbroken regime and the broken regime display distinctive behaviors for the non-Hermitian AA model with $\mathcal{PT}$ symmetry. We also consider the evolution of the mean displacement $\sigma(t)$ for different strengths of the hopping term $J$ when $\lambda$ is set as the unit energy shown in Appendix A. It displays similar results to those discussed above.
	
	Figure \ref{Fig31} shows the time evolution of the mean displacement $\sigma(t)$ with different lattice lengths $L$ and $\lambda$ to show the size-induced physics in our system. As shown in Fig. \ref{Fig31}(a) with $\lambda=0.8$ in the extended regime, both cases with $L=F_{13}=377$ and $L=F_{16}=1597$ exhibit the same dynamic feature, \emph{i.e.}, the diffusion exponent $\tilde{\gamma}=1$. For the case with $\lambda=1$, $\sigma(t) \sim t$ independent of the lattice length $L$ shown in Fig. \ref{Fig31}(b). When the parameter $\lambda$ enters into the $\mathcal{PT}$ symmetry broken localization regime, $\sigma(t)$ for $L=377$ and $1597$ with the initial wave packet localized at the center of the lattice shows similar dynamical features to those shown in Figs. \ref{Fig31}(c) and \ref{Fig31}(d). As seen in Figs. \ref{Fig31}(c) with $\lambda=1.5$ ($\tilde{\gamma}\approx 2$) and \ref{Fig31}(d) with $\lambda=3$ ($\tilde{\gamma}\approx 5$), a remarkable point is that the $\lambda$-dependent diffusion exponent $\tilde{\gamma}>1$ of the hyperdiffusion process increases with the increase of the non-Hermitian modulation strength. In Appendix B, we also discuss the cases of the chain's length deviating from a Fibonacci number. When the wave packet spreads near the boundaries, the evolution of the initially localized wave packet with the parameter $\lambda$ in the extended and multifractal regimes brings about a sudden rise in a short time, and then the wave packet is frozen at the boundaries, which is different from the one with the chain's length being a Fibonacci number. When $\lambda$ is localized in the $\mathcal{PT}$ symmetry broken regime, the evolution process is $L$-dependent. Investigations on the size-dependent features in quasi-periodic chains have been reported in Refs. \cite{Znidaric1,Varma,Znidaric2,Purkayastha,Znidaric3,Purkayastha1}. Our results imply that for the case with a chain's length being a Fibonacci number, the dynamical evolution for different $L$ shows a similar feature.
	
	\section{Loschmidt echo dynamics}
	
	The Loschmidt echo plays a significant role in characterizing the dynamical signature of the quantum phase transition. It was shown that the Loschmidt echo evolution could characterize the localization-delocalization transition in a Hermitian AA model \cite{Zhihao2,YangChao}. When the initial and post-quench systems are located in the same localization regime, the Loschmidt echo will oscillate without decaying to zero in a long time. However, if they are located in different localization regimes, the Loschmidt echo will decay and reach near $0$ at some time intervals. However, study of the Loschmidt echo dynamics for a $\mathcal{PT}$ symmetric AA model has not been demonstrated.
	
	In this section, we focus on the quench dynamics of the non-Hermitian AA model described by the Hamiltonian (\ref{eq1}). The system is initially prepared in an eigenstate $|\Psi(t_0)\rangle$ of the Hamiltonian $\hat{H}(\lambda^i)$ at time $t_0$ with $\langle \Psi(t_0)|\Psi(t_0)\rangle=1$, and then suddenly quenched to the final Hamiltonian $\hat{H}(\lambda^f)$. We define the return amplitude
	\begin{equation}\label{eq6}
		G(t;\lambda^i,\lambda^f)=\langle \Psi(t_0)|\Psi(t)\rangle
	\end{equation}
	where
	\begin{equation}\label{eq7}
		|\Psi(t)\rangle=\frac{1}{\sqrt{\mathcal{N}}}e^{-i(t-t_0)\hat{H}(\lambda^f)}|\Psi(t_0)\rangle,
	\end{equation}
	with $\mathcal{N}$ being the normalization coefficient of the time-evolution wave function $|\Psi(t)\rangle$, and setting $t_0=0$. The behavior of the return probability (Loschmidt echo) can be described by
	\begin{equation}\label{eq8}
		\mathcal{L}(t;\lambda^i,\lambda^f)=|G(t;\lambda^i,\lambda^f)|^2,
	\end{equation}
	where the superscripts $i$ and $f$ correspond to before and after the quench process, respectively. The distinctive dynamics in different $\mathcal{PT}$ symmetric regimes suggest us the we should study the quench processes in $\lambda^f \le J$ and $\lambda^f >J$, respectively.
	
	\subsection{Quench processes with $\lambda^f \le J$}
	
	\begin{figure}[tbp]
		\begin{center}
			\includegraphics[width=.55 \textwidth] {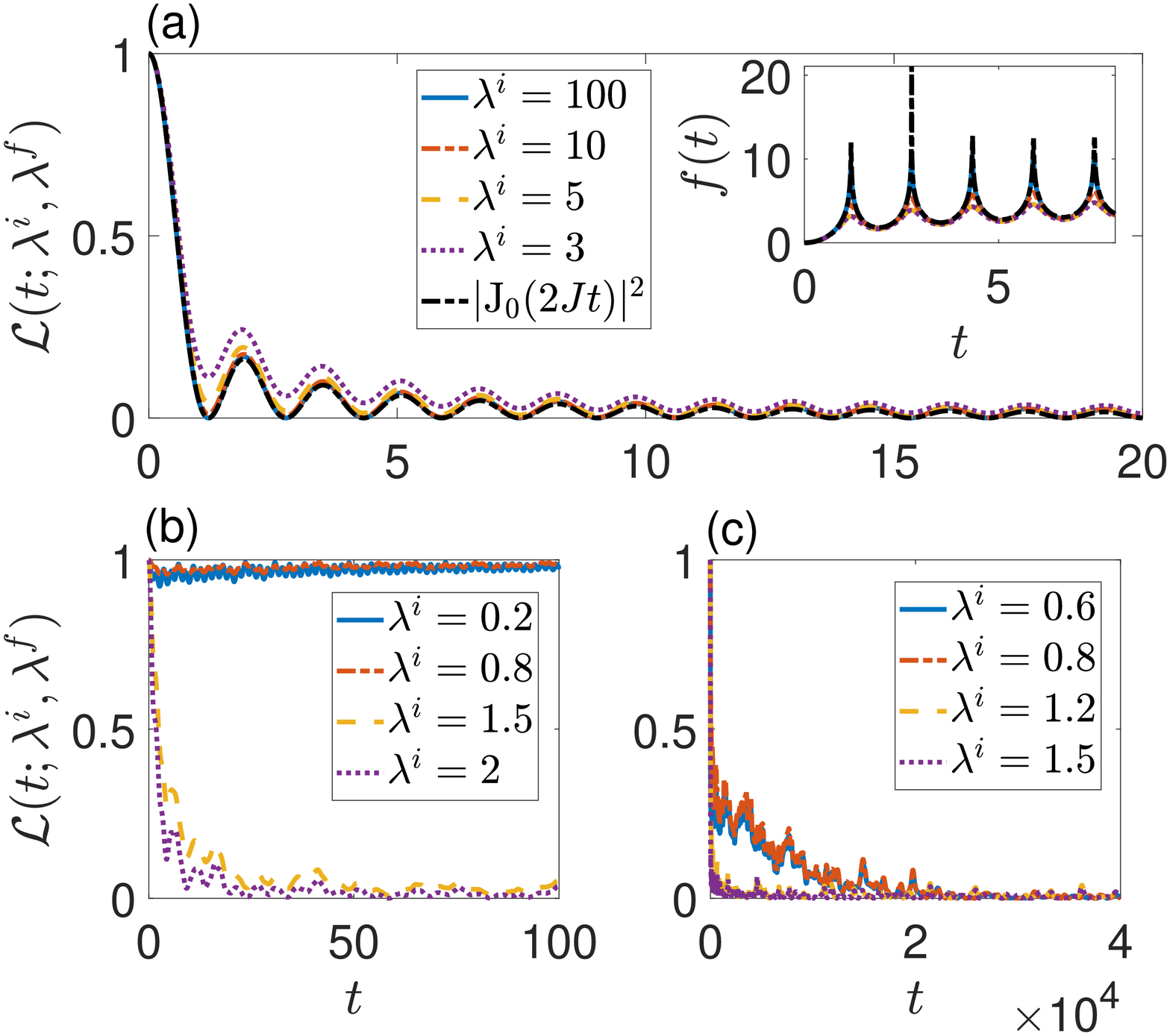}
		\end{center}
		\caption{(Color online) Evolution of the Loschmidt echo. The initial state is chosen to be the eigenstate of the energy with the lowest real part at different pre-quenched modulation amplitude $\lambda^i$, and the final Hamiltonian with the parameter (a) $\lambda^f=0$, (b) $\lambda^f=0.6$, and (c) $\lambda^f=1$, respectively. The inset of (a) shows the evolution of dynamical free energy $f(t)$ for different $\lambda^i$ and $\lambda^f=0$. Here, $J=1$ and $L=610$. }\label{Fig4}
	\end{figure}
	
	\begin{figure}[tbp]
		\begin{center}
			\includegraphics[width=.45 \textwidth] {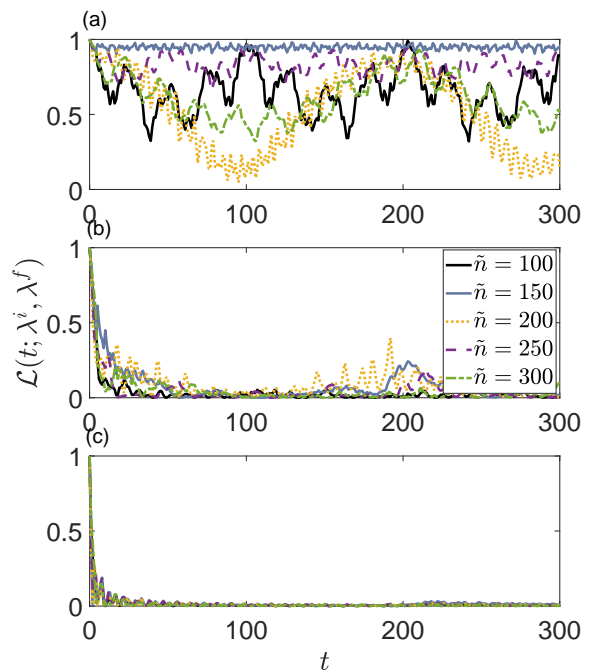}
		\end{center}
		\caption{(Color online) Evolution of the Loschmidt echo with the final modulation amplitude $\lambda^f=0.6$. The initial states are chosen to be the given eigenstates with the energies $E_{\tilde{n}}$ for (a) $\lambda^i=0.8$, (b) $\lambda^i=1.0$, and (c) $\lambda^i=1.5$, respectively.  The real part of the energies is ordered in ascending order.  
		Here, $J=1$ and $L=377$. }\label{Fig5}
	\end{figure}
	
	We first consider the case of the parameter after the quench process localized in the $\lambda^f\le J$ region, which is a $\mathcal{PT}$ symmetry unbroken regime with all energy spectra being real. For the limiting case of the quench process from $\lambda^i\to \infty$ to $\lambda^f=0$, the initial state is localized in a single site, $|\psi_m^{\lambda^i\to\infty}\rangle=\sum_j \delta_{jm}\hat{c}_j^{\dagger}|0\rangle=|m\rangle$, and the corresponding eigenenergy $E_m=\lambda^i\exp{(-i2\pi\alpha m)}$. By performing a quench process to $\lambda^f=0$, the eigenstates for the case of $\lambda^f=0$ can be written as $|\psi_k^{\lambda^f=0}\rangle=1/\sqrt{L}\sum_{j}e^{ikj}\hat{c}_j^{\dagger}|0\rangle=|k\rangle$, with the wave vector $k=2\pi l/L$ ($l=1,2,\cdots,L$), and the corresponding eigenvalue is $E_k=2J\cos{k}$.
	The return amplitude is 
	\begin{align}\label{eq9}
		G(t;\infty,0)&=\langle m|e^{-iH(\lambda^f)t}|m\rangle =\sum_k \langle m|e^{-iH(\lambda^f)t}|k\rangle\langle k|m\rangle \notag \\
		&=\frac{1}{L}\sum_k e^{-i2Jt\cos{(k)}}.
	\end{align}
	In the thermodynamic limit ($L\to \infty$), $G(t)=\mathrm{J}_0(2Jt)$ with $\mathrm{J}_0(x)$ being the zero-order Bessel function of the first kind. It is clear that the zeros of the Loschmidt echo $t^*$ occur at the halfway point of the zeros of $\mathrm{J}_0(x)$. The emergence of the zeros of the Loschmidt echo implies that the initial and final states are localized in different localized regimes, according to previous results in the Hermitian cases. Figure \ref{Fig4}(a) shows the evolution of the Loschmidt echo with $\lambda^f=0$ and different $\lambda^i>J$. In this case, the Loschmidt echoes $\mathcal{L}(t)$ for $\lambda^i=100$, $10$, $5$, and $3$ oscillate at the same frequency as the analytical result. This implies that the frequencies of the Loschmidt echo are not sensitive to the initial parameter $\lambda^i$ as long as $\lambda^i$ is large enough. To see the zeros of $\mathcal{L}(t)$ more clearly, we introduce the dynamical free energy $f(t)=-\ln{\mathcal{L}(t)}$, which is shown in the inset of Fig. \ref{Fig4}(a). It will be divergent at the dynamical phase transition time $t^*$. In the large $\lambda^i$ limit, $f(t)$ exhibits obvious peaks at $t=t^*$ that almost completely overlap the analytical result. With the decrease of $\lambda^i$, the change of the peak positions of $f(t)$ is tiny, while the peak amplitudes of $f(t)$ decrease.
	
	The analytical result for the limit case of the quench process from $\lambda^i=\infty$ to $\lambda^f=0$ is close to the $0$ of the Loschmidt echo. Now, we consider the general cases in which $\lambda^i$ and $\lambda^f$ deviate from the limit case for $\lambda^f\le J$. Figures \ref{Fig4}(b) and \ref{Fig4}(c) show the evolution of the Loschmidt echo with different $\lambda^i$, the parameter of the final Hamiltonian deviating from zero, and the initial state being chosen to be the eigenstate of the energy with the lowest real part at different $\lambda^i$. In Fig. \ref{Fig4}(b), we choose $\lambda^f=0.6$, where all the eigenstates are extended. For $\lambda^i=0.2$ and $0.8$, $\mathcal{L}(t)$ oscillates and has a positive lower bound that never approaches $0$ during the evolution. However, if $\lambda^i>J$, $\mathcal{L}(t)$ approaches $0$ after some time intervals [see Fig. \ref{Fig4}(b) for $\lambda^f=1.5$ and $2$]. Figure \ref{Fig4}(c) shows the dynamics of the Loschmidt echo with $\lambda^f=1$ where the system is localized in the multifractal phase. As seen in Fig. \ref{Fig4}(c), the long-time evolution of the Loschmidt echo approaches $0$ for the initial states localized in either the extended ($\lambda^i=0.6$ and $0.8$) or the localized ($\lambda^i=1.2$ and $1.5$) regime. 
	
	To show that our result is independent of the initial eigenstate's choice, we consider the different eigenstates as the initial states to calculate the evolution of the Loschmidt echo. As a concrete example, we choose $\lambda^f=0.6$, and the initial states are the $\tilde{n}$-th eigenstates of the corresponding eigenenergies $E_{\tilde{n}}$ with $\tilde{n}=100$, $150$, $200$, $250$, and $300$ as shown in Fig. \ref{Fig5}. The real part of the $E_{\tilde{n}}$ is ordered in ascending order. For the case of $\lambda^i=0.8$ in the extended regime shown in Fig. \ref{Fig5}(a), the Loschmidt echoes for different initial states oscillate and never approach $0$. However, for $\lambda^i=1.0$ and $1.5$ localized in different regimes from the one with $\lambda^f=0.6$ [see Figs. \ref{Fig5}(b) and \ref{Fig5}(c), respectively], the dynamics of the Loschmidt echo for different initial eigenstates decay in an oscillating way and can always touch near $0$ in some time intervals.
	
	In conclusion, for the case of the parameter of the post-quench Hamiltonian localized in the $\lambda \le J$ region where all the eigenenergies are real, the behaviors of the Loschmidt echo are similar to the Hermitian cases in that the dynamical signature of the localized transition can be characterized by the emergence of zero points in the evolution of the Loschmidt echo.
	
	\subsection{Quench processes with $\lambda^f>J$}
	
	\begin{figure}[tbp]
		\begin{center}
			\includegraphics[width=.5 \textwidth] {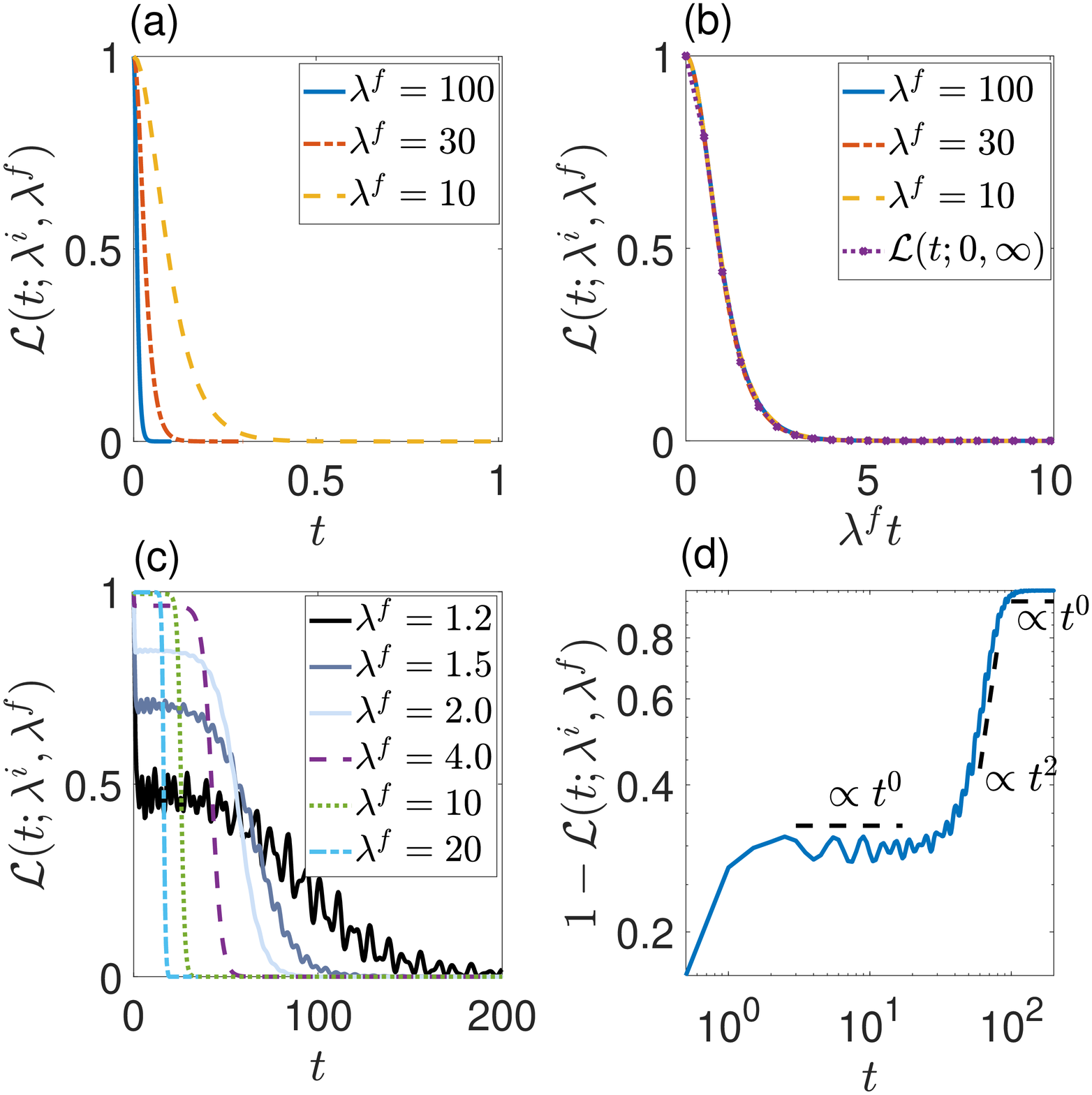}
		\end{center}
		\caption{(Color online) (a) and (b) Evolution of the Loschmidt echo for the large final modulation amplitude $\lambda^f$ with $L=610$, and the initial state being chosen to be the ground state at $\lambda^i=0$. (a) $\mathcal{L}(t;0,\lambda^f)$ versus $t$. (b) $\mathcal{L}(t;0,\lambda^f)$ versus the rescaled time $\lambda^f t$. (c) Evolution of the Loschmidt echo for finite $\lambda^f$ with $L=377$, and the initial state being chosen to be the wave packet localized at the center of the lattice. (d) Log-log plot of $1-\mathcal{L}(t;\lambda^i,\lambda^f)$ as a function of time $t$ for $\lambda^f=1.5$ with $L=377$, and the initial state being chosen to be the wave packet localized at the center of the lattice. The black dashed line represents a power-law fitting. Here, $J=1$.}\label{Fig6}
	\end{figure}

	In this subsection, we consider the case of the final Hamiltonian with the parameter $\lambda^f>J$, where all the eigenstates are localized and the $\mathcal{PT}$ symmetry is broken. First, an analytical calculation is considered for the quench process with $\lambda^i=0$ and $\lambda^f=\infty$. For $\lambda^i=0$, the system is initially prepared in a plane-wave state $|k\rangle$ with the eigenvalue $E_k=2J\cos{k}$. By performing a sudden quench to $\hat{H}(\lambda^f)$ in the limit of $\lambda^f\to \infty$, the eigenstates of $\hat{H}(\lambda^f)$ are localized in a single site $|m\rangle$, and the corresponding eigenenergy is $E_m=\lambda^f e^{-i 2\pi\alpha m}$. The evolution wave function can be written as
	\begin{align}\label{eq10}
		|\Psi(t;\lambda^i=0,\lambda^f\to \infty)\rangle &=\frac{1}{\sqrt{\mathcal{N}}}e^{-i\hat{H}(\lambda^f)t}|k\rangle \notag \\
		&= \frac{1}{\sqrt{\mathcal{N}L}}\sum_m e^{ikm}e^{-i\lambda^f t e^{-i2\pi\alpha m}}|m\rangle.
	\end{align}
	In the large $L\to \infty$ limit, we can obtain
	\begin{equation}\label{eq11}
		G(t;0,\infty)=\frac{1}{\sqrt{\mathcal{N}}}\sum_{\nu=-\infty}^{\infty} (-1)^{\nu} \mathrm{J}_{\nu}(\lambda^f t)\mathrm{I}_{\nu}(\lambda^f t),
	\end{equation}
	where $\mathcal{N}=\mathrm{I}_0(2\lambda^f t)$, with 
	$\mathrm{J}_{\nu}(x)$ being the Bessel function of the first kind of the index $\nu$, and $\mathrm{I}_{\nu}(x)$ being the modified Bessel function of the first kind of $\nu$ order. Figures \ref{Fig6}(a) and \ref{Fig6}(b) show the evolution of the Loschmidt echo with the initial state chosen to be the ground state of $\hat{H}(\lambda^i=0)$ and large $\lambda^f$ as a function of $t$, and the rescaled time $\lambda^f t$, respectively. The Loschmidt echoes for $\lambda^f=10$, $30$, and $100$ rapidly decay to $0$ with time, and the larger $\lambda^f$ is, the faster $\mathcal{L}(t)$ decays. They nearly overlap with the analytical result $\mathcal{L}(t;0,\infty)=1/\mathcal{N}|\sum_{\nu} (-1)^{\nu}\mathrm{J}_{\nu}(\lambda^f t)\mathrm{I}_{\nu}(\lambda^f t)|^2$ by rescaling the time $\lambda^f t$ shown in Fig. \ref{Fig6}(b). 
	
	Another limit is also considered, i.e., the parameter $\lambda^i \to \infty$ of the Hamiltonian (\ref{eq1}) quenched to finite $\lambda^f > J$. As an example, we set the initial state to be the wave packet localized at the center of the lattice $|j_0\rangle$, which is shown in Figs. \ref{Fig6}(c) and \ref{Fig6}(d). It is known that in the $\lambda^f\to\infty$ limit,  $\mathcal{L}(t;\infty,\infty)=1$, due to the initial state being the eigenstate of the post-quench Hamiltonian. When $\lambda^f$ deviates from the infinite value, the initial state $|j_0\rangle$ is no longer the eigenstate of $\hat{H}(\lambda^f)$, and the excited single-channel has a superposition of various eigenstates. No matter how small the superposition with the initial condition is, the mode with the largest imaginary part among them will dominate after a finite evolution time, and a non-Hermitian jump process occurs that will induce an evident change of the return amplitude. Hence, for a finite $\lambda^f$, the Loschmidt echo no longer remains uniform, and it will present complex features. As seen in Fig. \ref{Fig6}(c), the Loschmidt echo first decays to a finite value $\mathcal{L}_{c}$ in a short time $\tilde{t}_1$ corresponding to a finite overlap between the evolution wave function $|\Psi(t_1)\rangle$ and $|j_0\rangle$.  $\mathcal{L}(t;\lambda^i,\lambda^f)$ then displays an oscillation around the limited value $\mathcal{L}_{c}$ in a certain time interval $\Delta\tilde{t}$. $\mathcal{L}_{c}$ and $\Delta\tilde{t}$ depend on $\lambda^f$. In the large-$\lambda^f$ limit, the value of $\mathcal{L}_{c}$ tends to $1$ and $\Delta\tilde{t}\to 0$. With the decrease of $\lambda^f$, $\mathcal{L}_{c}$ decreases, while $\Delta\tilde{t}$ increases in the large-$\lambda^f$ case and decreases for a finite $\lambda^f$. Finally, after the temporary localization process, the Loschmidt echo evolution displays a $\lambda^f$-dependent damping to $0$, which shows that the overlap of the final state and $|j_0\rangle$ tends to $0$, and it corresponds to the emergence of a non-Hermitian jump process. Figure \ref{Fig6}(d) shows the log-log plot of $1-\mathcal{L}(t;\lambda^i,\lambda^f)$ as the function of time $t$ for $\lambda^f=1.5$. We can see that the Loschmidt echo presents similar behaviors to the mean displacement $\sigma(t)$ in this limit, and the same power-law indices are found in the corresponding evolution stages. 
	
	\begin{figure}[tbp]
		\begin{center}
			\includegraphics[width=.5 \textwidth] {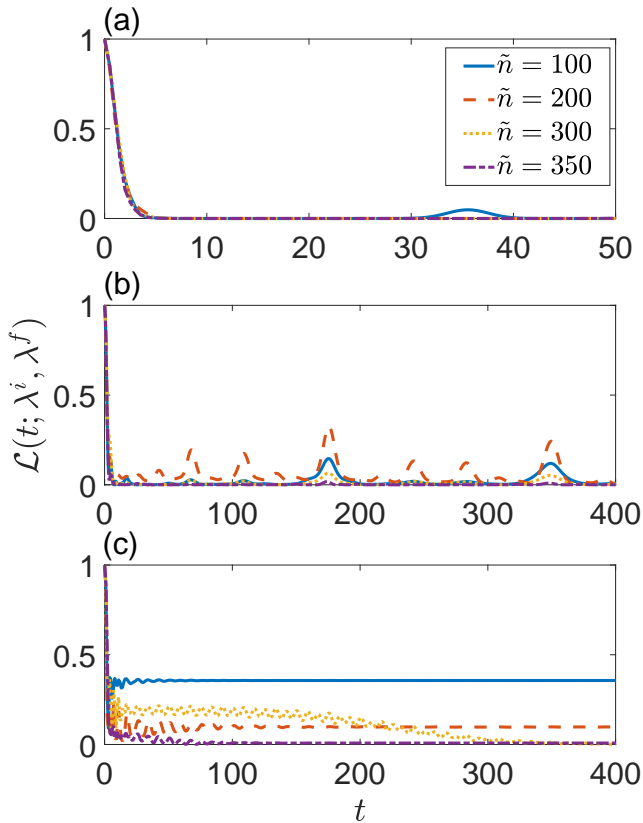}
		\end{center}
		\caption{(Color online) Evolution of the Loschmidt echo with $\lambda^f=1.5$. The initial states are chosen to be the given eigenstates with the energies $E_{\tilde{n}}$ for (a) $\lambda^i=0.6$, (b) $\lambda^i=1.0$, and (c) $\lambda^i=1.2$, respectively. The energies $E_{\tilde{n}}$ have been ordered according to increasing values of the real part. Here, $J=1$ and $L=377$.}\label{Fig7}
	\end{figure}
	
	The results of two limits suggest that the existence of zero points during the Loschmidt echo evolution seems to no longer be a dynamical signature of the localization-delocalization transition for the case of $\lambda^f >J$. To clarify such a conjecture, we calculate the evolution of the Loschmidt echo with $\lambda^f=1.5$ for different $\lambda^i$ and $\tilde{n}$, with the real part of the energies in ascending order, which is shown in Fig. \ref{Fig7}. When $\lambda^i$ is localized in the extended ($\lambda^i=0.6$) or the multifractal regime ($\lambda^i=1.0$), the evolution of the Loschmidt echo exhibits a series of zero points for different $\lambda^i$ and $\tilde{n}$. However, for the initial state being an eigenstate of $\hat{H}(\lambda^i)$ with $\lambda^i>J$, e.g., $\lambda^i=1.2$ shown in Fig. \ref{Fig7}(c), the Loschmidt echo is strongly dependent on the initial state, and the conclusion of the Hermitian case is no longer available for the non-Hermitian one in the $\mathcal{PT}$ symmetry unbroken regime.  
	
	\section{Conclusion}
	In this paper, we study the dynamics evolution of the non-Hermitian AA model with $\mathcal{PT}$ symmetry. The $\mathcal{PT}$ symmetric AA model exhibits a $\mathcal{PT}$ symmetry broken point at $\lambda=J$. When $\lambda<J$, all the eigenenergies are real and the corresponding eigenstates are extended, while for $\lambda>J$, the complex energies emerges and the corresponding states are localized. The states at the transition point are multifractal. We can apply the evolution of the profile of the wave function and the long-time survival probability to distinguish the localization properties of the system. The evolution of the mean displacement displays distinctive behaviors in different $\mathcal{PT}$ symmetric regimes, which is also available for the evolution of the Loschmidt echo. According to our calculation, when the post-quench parameter is localized in the $\mathcal{PT}$ symmetry unbroken regime, the behaviors of the Loschmidt echo are similar to the Hermitian cases in that the dynamical signature of the localization-delocalization transition can be characterized by the emergence of a series of zero points in the evolution of the Loschmidt echo. However, when the post-quench parameter is localized in the $\mathcal{PT}$ symmetry broken regime, the dynamical detection of the transition point by the Loschmidt echo method is unavailable. Similar conclusions for the Loschmidt echo in some other $\mathcal{PT}$ symmetric systems without disorder have been reported \cite{Znojil,Krejcirik,KWang}.  Our results can be easily examined in the silicon waveguide platform, cold atomic gases, and ion chains \cite{Bylinskii1,Bylinskii2,Bonetti,Benassi,Mandelli,Kiethe}.
	
	\begin{acknowledgements}
		Z. Xu is supported by the NSFC (Grants No. 11604188 and No. 12047571), Beijing National Laboratory for Condensed Matter Physics, and STIP of Higher Education Institutions in Shanxi under Grant No. 2019L0097. S. Chen is supported by the National Key Research and Development Program of China (2016YFA0300600 and 2016YFA0302104), NSFC under Grants No.11974413, and the Strategic Priority Research Program of Chinese Academy of Sciences under Grant No. XDB33000000. This work is also supported by NSF for Shanxi Province Grant No.1331KSC.
		
	\end{acknowledgements}

	\section*{Appendix A: Effects of $J$ with fixed $\lambda$}
	\begin{figure}[tbp]
		\begin{center}
			\includegraphics[width=.5 \textwidth] {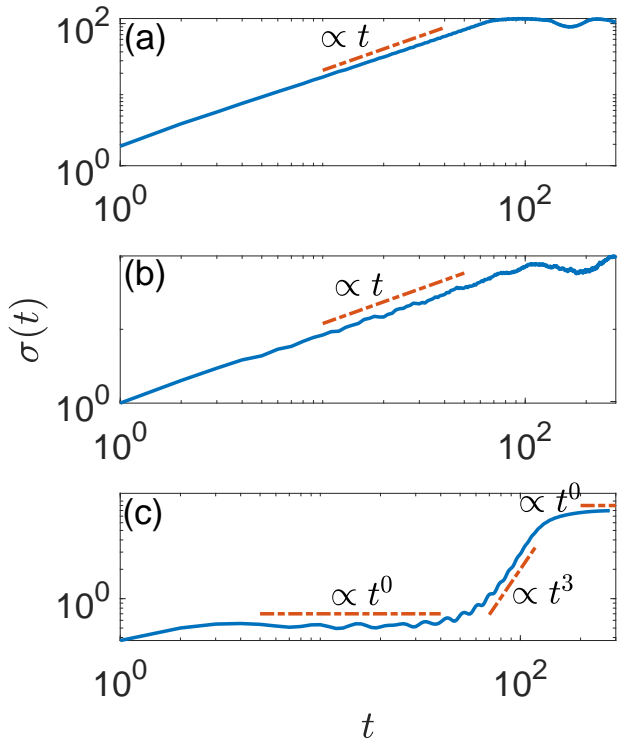}
		\end{center}
		\caption{(Color online) Time evolution of the mean displacement $\sigma(t)$ with the hopping strength (a) $J=1.5$, (b) $J=1.0$, and (c) $J=0.5$, respectively. Here, $\lambda=1$ and $L=377$.}\label{SFig1}
	\end{figure}
	In this appendix, we consider $\lambda$ as the unit energy, and the evolution of the mean displacement $\sigma(t)$ with different strengths of the hopping term is shown in Fig. \ref{SFig1}. In this case, when $J>1$, the system is localized in the extended regime corresponding to a ballistic diffusion with $\tilde{\gamma}=1$ seen in Fig. \ref{SFig1}(a). For the multifractal case [Fig. \ref{SFig1}(b)], which is also studied in the main text, it displays the diffusion exponent $\tilde{\gamma}=1$. When the system is in the localization regime [Fig. \ref{SFig1}(c) with $J=0.5$], the similar structure of $\sigma(t)$ is detected as shown in main text with $J=1$ and $\lambda=1.5$, though the diffusion exponent $\tilde{\gamma}\approx 3$ in the hyperdiffusion stage. For the case in the $\mathcal{PT}$ symmetry broken regime, $\tilde{\gamma}$ is dependent on $\lambda/J$ for a hyperdiffusion.

	\section*{Appendix B: Chain's length deviating from a Fibonacci number}
	\begin{figure}[tbp]
		\begin{center}
			\includegraphics[width=.5 \textwidth] {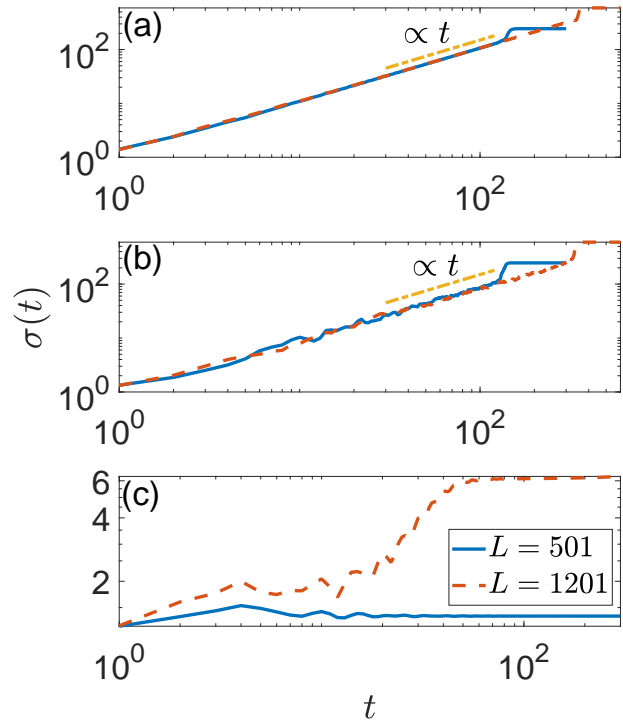}
		\end{center}
		\caption{(Color online) Time evolution of the mean displacement $\sigma(t)$ with different $L$ and the non-Hermitian modulation strength (a) $\lambda=0.6$, (b) $\lambda=1.0$, and (c) $\lambda=1.5$, respectively. Here, $J=1$.}\label{SFig2}
	\end{figure}

	\begin{figure}[tbp]
		\begin{center}
			\includegraphics[width=.5 \textwidth] {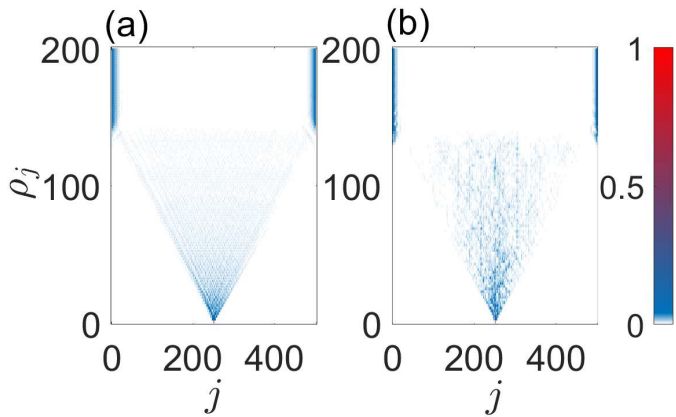}
		\end{center}
			\caption{(Color online)  Evolutions of the profile of the wave function $\rho_j(t)$ with the non-Hermitian modulation amplitude (a) $\lambda=0.6$ and (b) $\lambda=1.0$, respectively. Here, $J=1$ and $L=501$.}\label{SFig3}
	\end{figure}
	In this appendix, we consider the dynamics of the non-Hermitian AA model described by the Hamiltonian (\ref{eq1}) and (\ref{seq1}) in the main text with the chain's length deviating from a Fibonacci number. This is another useful direction to study the transport in quasi-periodic potentials \cite{Znidaric1,Varma,Znidaric2,Purkayastha,Znidaric3,Purkayastha1}. Figure \ref{SFig2} shows the time evolution of the mean displacement $\sigma(t)$ with the chain's length $L=501$ and $1201$. The initial wave packet is localized at the center of the lattice. As shown in Figs. \ref{SFig2} (a) and \ref{SFig2}(b) with $\lambda=0.6$ and $1.0$, respectively, we can find that before the wave packet spreading near the boundaries, $\sigma(t) \propto t$. When the evolution of the wave packet is near the boundaries, the mean displacement brings about a sudden rise to a finite value in a short time. To intuitively see the phenomenon displaying in the extended and multifractal regimes, we show the evolution of the profile of the wave packet $\rho_j(t)$ in Fig. \ref{SFig3}. The evolution of the packet in both regimes exhibits similar features to the chain's length being a Fibonacci number before the wave packet reaching the boundaries. Then the wave packet is frozen at the boundaries corresponding to a stationary value of $\sigma(t)$.  When the localized wave packet spreads in the $\mathcal{PT}$ symmetry broken regime, $\sigma(t)$ exhibits an $L$-dependent feature, which is shown in Fig. \ref{SFig2}(c).

\end{document}